\documentclass[prd,twocolumn,nofootinbib,amsfonts,amssymb]{revtex4}

\usepackage{graphicx}

\usepackage{bm}

\newcommand{\beq}{\begin{equation}}
\newcommand{\eeq}{\end{equation}}
\newcommand{\beqs}{\begin{eqnarray}}
\newcommand{\eeqs}{\end{eqnarray}}

\begin{document}

\title{$\beta'_{IR}$ at an Infrared Fixed Point in Chiral Gauge Theories}

\author{Thomas A. Ryttov$^a$ and Robert Shrock$^b$}

\affiliation{(a) \ CP$^3$-Origins, University of Southern Denmark, 
 Campusvej 55, Odense, Denmark}

\affiliation{(b) \ C. N. Yang Institute for Theoretical Physics and
Department of Physics and Astronomy, \\
Stony Brook University, Stony Brook, NY 11794, USA }

\begin{abstract}

  We present scheme-independent calculations of the derivative of the beta
  function, denoted $\beta'_{IR}$, at a conformally invariant infrared (IR)
  fixed point, in several asymptotically free chiral gauge theories, namely
  SO($4k+2$) with $2 \le k \le 4$ with respective numbers $N_f$ of fermions in
  the spinor representation, and E$_6$ with fermions in the fundamental
  representation.

\bigskip


\end{abstract}

\maketitle

% =======================================================================

\section{Introduction}
\label{intro_section}

The properties of an asymptotically free gauge theory at an infrared fixed
point (IRFP) of the renormalization group (RG) in a conformally invariant
regime are of fundamental interest. Owing to the asymptotic freedom, one can
perform reliable perturbative calculations in the deep ultraviolet (UV) where
the gauge coupling approaches zero, and then follow the RG flow to the
infrared.  This flow is described by the beta function, $\beta =
d\alpha/d\ln\mu$, where $\alpha=g^2/(4\pi)$, $g=g(\mu)$ is the running gauge
coupling, and $\mu$ is a Euclidean momentum scale. The value of $\alpha$ at the
IRFP is denoted $\alpha_{IR}$. For a given gauge group $G$ and fermion
representation $R$ of $G$, the requirement of asymptotic freedom places an
upper bound on the number of fermions transforming according to this
representation.  If the number of these fermions is slightly less than the
maximum allowed by asymptotic freedom, then the theory flows from the UV to a
weakly coupled IRFP in a non-Abelian Coulomb phase.  At this IRFP the theory is
scale-invariant and is inferred to be conformally invariant \cite{scalecon}.
Physical quantities at this IRFP can be expressed perturbatively as series
expansions in powers of $\alpha_{IR}$ (e.g., \cite{bvh,ps,flir}).  However,
beyond respective low loop orders, the coefficients in these expansions depend
on the scheme used for regularization and renormalization of the theory.
Consider an asymptotically free vectorial gauge theory (VGT) with gauge group
$G$ and $N_f$ Dirac fermions in a representation $R$ of $G$, such that the RG
flow leads to an IRFP, and let $N_u$ denote the value of $N_f$ such that for
$N_f < N_u$, the theory is asymptotically free. Since $\alpha_{IR}$ becomes
small as $N_f$ approaches $N_u$ from below, one can reexpress physical
quantities as series expansions in the manifestly scheme-independent variable
$N_u-N_f$ \cite{bz}.  This has the advantage that the coefficients in the
expansion are scheme-independent.  

Recently, for vectorial gauge theories, we have calculated such
scheme-independent expansions in \cite{gtr}-\cite{dexo} for anomalous
dimensions of (gauge-invariant) fermion bilinears and the derivative of the
beta function, evaluated at the IRFP,
\beq
\frac{d\beta}{d\alpha} \Big |_{\alpha=\alpha_{IR}} \equiv \beta'_{IR} \ .
\label{betaprime_def}
\eeq
These are both physical quantities and hence are scheme-independent
\cite{gross75}. The derivative $\beta'_{IR}$ is equivalent to the anomalous
dimension of ${\rm Tr}(F_{\mu\nu}F^{\mu\nu})$, where $F_{\mu\nu}^a$ is the
field-strength tensor \cite{traceanomaly}.  

Here we extend our analysis to asymptotically free chiral gauge theories,
denoted $\chi$GTs (in four spacetime dimensions).  We consider several such
theories, which can be classified into two general types: (i) theories with
special orthogonal gauge groups $G={\rm SO}(N)$, where $N=4k+2$ with $k \ge 2$,
containing $N_f$ chiral fermions transforming according to the spinor
representation $S$ of this group; and (ii) theories with the
exceptional gauge group $G={\rm E}_6$, containing $N_f$ chiral fermions in the
fundamental (27-dimensional) representation.  These representations are complex
\cite{mehta}. Without loss of generality, all
fermions may be taken as left-handed. We present scheme-independent
calculations of $\beta'_{IR}$ to $O(\Delta_f^5)$ in these theories, where
\beq
\Delta_f = N_u - N_f \ .  
\label{deltaf}
\eeq
The fermions are massless, since fermion mass terms are forbidden by the chiral
gauge invariance.  For the same reason, the anomalous dimensions
$\gamma_{\bar\psi\psi,IR}$ are gauge-dependent here and hence are not of
physical interest, so we focus on $\beta'_{IR}$.  As will be shown in Section
\ref{methods_section}, the constraint of asymptotic freedom limits our
consideration of SO($4k+2$) theories to those with $k=2, \ 3, \ 4$, i.e.,
SO(10), SO(14), and SO(18), and, for each of these, this constraint yields
corresponding upper limits on $N_f$, namely $N_f \le 21$ for SO(10), $N_f \le
8$ for SO(14), and $N_f \le 2$ for SO(18). Similarly, the requirement of
asymptotic freedom yields the upper limit $N_f \le 21$ in the E$_6$ theory. Our
SO($4k+2$) theories with $k \ge 2$ and our E$_6$ chiral gauge theory have no
gauge anomaly \cite{gg,gp} and no global Witten-type anomaly \cite{witten},
since the relevant $\pi_4$ homotopy groups are trivial \cite{pi4}.  (Theories
that have complex representations and vanishing gauge anomaly are commonly
called safe.)  There is currently renewed interest in four-dimensional
conformal field theories (some reviews include \cite{cft}).  Our results serve
as useful input to the further study of conformally invariant gauge theories.

% ========================================================================

\section{Background and Methods} 
\label{methods_section} 

% ----------------------------------------------------------------------

\subsection{General}

Here we briefly review some background and methods relevant for our work.  We
refer the reader to our previous papers \cite{gtr}-\cite{dexo} for details. As
noted, we focus on chiral gauge theories that have complex representations
but for which the gauge anomaly vanishes identically.  These include theories
with the gauge groups $G={\rm SO}(N)$, where $N=4k+2$ with $k \ge 2$ and
$G={\rm E}_6$ \cite{gg,gp}.  The requirement of asymptotic freedom implies that
$N_f$ must be less than an upper ($u$) bound $N_u$, where
\beq
N_u = \frac{11C_A}{2T_f} \ .
\label{nfb1z}
\eeq
Here, $C_2(R)$ is the quadratic Casimir invariant for the representation $R$,
$C_A=C_2(A)$, where $A$ is the adjoint representation, and 
$T_f \equiv T(R)$ is the trace invariant (see appendix).

For a chiral gauge theory, the two-loop ($2\ell$) beta function (which is
scheme-independent) has an IR zero if $N_f$ lies in the interval $I$ defined by
$N_{\ell} < N_f < N_u$, where \footnote{
Here and elsewhere, when an expression is given for $N_f$
that formally evaluates to a non-integral real value, it is understood
implicitly that one infers an appropriate integral value from it.} 
\beq
N_{\ell} = \frac{17C_A^2}{T_f(5C_A+3C_f)} \ .
\label{nfb2z}
\eeq
This IR zero occurs at
\beq
\alpha_{IR,2\ell} = \frac{2\pi (11C_A-2T_fN_f)}{T_f(5C_A+3C_f)N_f-17C_A^2}  \ .
\label{alfir_2loop}
\eeq
Formally generalizing $N_f$ from positive integers ${\mathbb N}_+$ to positive
real numbers ${\mathbb R}_+$, one can let $N_f$ approach $N_u$ from below,
thereby making $\alpha_{IR,2\ell}$ arbitrarily small. Thus, for the UV to IR
evolution in this regime of $N_f$, one infers that the theory evolves from weak
coupling in the UV to an IRFP in a non-Abelian Coulomb phase (NACP, also called
conformal window).  We denote the lowest value of $N_f$ in this NACP as
$N_{f,cr}$. Our calculations assume that the IRFP is exact, as is the case in
the non-Abelian Coulomb phase \footnote{ For sufficiently small $N_f$,
  nonperturbative phenomena involving strong coupling can occur, including
  possible spontaneous chiral symmetry breaking with generic breaking of the
  chiral gauge symmetry, or, if 't Hooft anomaly-matching conditions are
  satisfied, confinement without global or gauge symmetry breaking. For a
  recent discussion and references to the literature, see, e.g., \cite{cgt}.
  We do not consider such phenomena at smaller $N_f$ here, instead restricting
  to the non-Abelian Coulomb phase, where the gauge and chiral symmetries are
  exact.}  $N_{f,cr} < N_f < N_u$.  In the analytic expressions and plots given
below, this assumption will be understood implicitly.

% ------------------------------------------------------------------------

\subsection{Interval $I$ for SO($4k+2$) Theories} 

For our SO($N$) theories with $N=4k+2$, $k \ge 2$, and chiral fermions in the
spinor representation $S$, one has $T_S = 2^{(N/2)-4}$ and 
$C_2(S) = N(N-1)/16$, so 
\beq
N_u = \frac{11(N-2)}{2^{(N/2)-3}} \ . 
\label{nu}
\eeq
Here $N_u$ takes on the values (i) 22 for $k=2$, i.e., SO(10); (ii) $33/4=8.25$
for $k=3$, i.e., SO(14); (iii) $11/4=2.75$ for $k=4$, i.e., SO(18); and (iv)
$55/64=0.859375$ for $k=5$, i.e., SO(22), decreasing monotonically toward zero
for larger $k$.  Hence, the only asymptotically free SO($4k+2$) chiral gauge
theories with chiral fermions in the spinor representation are as follows, for
physical integral $N_f$: (i) SO(10) with $1 \le N_f \le 21$; (ii) SO(14) with
$1 \le N_f \le 8$; and (iii) SO(18) with $N_f = 1, \ 2$. (The theories with
$N_f=0$ are pure gluonic theories and hence are not relevant here.)

For the SO($N$) gauge theories with $N=4k+2$, containing $N_f$ chiral fermions 
in the spinor representation, $N_\ell$ is given by 
\beqs
N_\ell &=& \frac{17(N-2)^2}{2^{(N/2)-8}(3N^2+77N-160)} \ . 
\label{nfb2z_so}
\eeqs
$N_\ell$ takes on the value (i) 4352/455 = 9.564835 for SO(10); (ii) 816/251 =
3.250996 for SO(14); and (iii) 1088/1099 = 0.9899909 for SO(18).  In Table
\ref{interval_so} we list the resultant intervals $I$ in $N_f$ for which the
asymptotically free chiral gauge theories of SO($4k+2$) type have a two-loop
beta function with an IR zero.  For each case, we give two ranges, namely one
for $N_f$ formally generalized to ${\mathbb R}_+$, and
the second for physical, integral $N_f \in {\mathbb N}_+$.

% -----------------------------------------------------------------------

\subsection{Interval $I$ for E$_6$ Theory} 

For the E$_6$ chiral gauge theory with $N_f$ fermions in the fundamental 
(27-dimensional) representation, $F$, $C_A \equiv C_2(G) = 12$, 
$T_F = 3$, and $C_2(F) = 26/3$, so $N_u = 22$. 
Hence, to maintain asymptotic freedom in this E$_6$ theory, we require that
$N_f < 22$.  Furthermore, we calculate that 
$N_{\ell} = 408/43 = 9.488372$. 
Therefore, the interval $I$ for this E$_6$ theory is 
\beqs
{\rm E}_6: && \ I: \quad 9.488 < N_f < 22 \ {\rm for} \ N_f \in {\mathbb R}_+,
\cr\cr
           && \ I: \quad 10 \le N_f \le 21 \ {\rm for} \ N_f \in {\mathbb N}_+
\label{interval_e6}
\eeqs
%

% -----------------------------------------------------------------------

\subsection{Scheme-Independent Expansion for $\beta'_{IR}$ } 

Given the property of asymptotic freedom, $\beta$ is negative in the
region $0 < \alpha < \alpha_{IR}$, and since $\beta$ is continuous, it follows
that this function passes through zero at $\alpha=\alpha_{IR}$ 
with positive slope, i.e., $\beta'_{IR} > 0$. This dervative $\beta'_{IR}$ 
has the scheme-invariant expansion
\beq
\beta'_{IR} = \sum_{j=2}^\infty d_j\Delta_f^{\ j} \ . 
\label{betaprime_delta_series}
\eeq
As indicated, $\beta'_{IR}$ has no term linear in $\Delta_f$.  In general, the
calculation of the scheme-independent coefficient $d_j$ requires, as inputs,
the $\ell$-loop coefficients in the beta function, $b_\ell$, for $1 \le \ell
\le j$.  For our calculation of $\beta'_{IR}$ to $O(\Delta_f^5)$ in
\cite{dexl}, we thus made use of the five-loop beta function from \cite{b5}.
In the literature, the beta function coefficients have usually been given for a
vectorial gauge theory with $N_f$ Dirac fermions in a representation $R$ of the
gauge group $G$. In the case of a chiral gauge theory with fermions in a single
representation of the gauge group, one can take over these results with the
replacement $N_f \to N_f/2$, reflecting the replacement of Dirac with chiral
fermions. In particular, we can use our previous calculations of the $d_j$ with
$2 \le j \le 4$ in \cite{dex} and $d_5$ in \cite{dexl} in a VGT for the
$\chi$GTs under consideration, with the correspondence
\beq
(d_j)_{\chi {\rm GT}} = 2^{-j} \, (d_j)_{\rm VGT} \ . 
\label{djrel_cgt_vgt}
\eeq

Let us denote the full scaling dimension of an operator ${\cal O}$ as $D_{\cal
  O}$ and its free-field value as $D_{{\cal O},free}$. We define the anomalous
dimension of ${\cal O}$, denoted $\gamma_{\cal O}$, by \footnote{
Some authors use the opposite sign convention for the anomalous dimension,
writing $D_{\cal O} = D_{\cal O,{\rm free}} + \gamma_{\cal O}$.} 
$D_{\cal O} = D_{{\cal O},free} - \gamma_{\cal O}$. Let the full scaling
dimension of ${\rm Tr}(F_{\mu\nu}F^{\mu\nu})$ be denoted $D_{F^2}$ (with
free-field value 4). At an IRFP, $D_{F^2,IR} = 4 + \beta'_{IR}$
\cite{traceanomaly}, so $\beta'_{IR} = -\gamma_{F^2,IR}$. Given that the theory
at an IRFP in the non-Abelian phase is conformally invariant, there is a
conformality bound from unitarity, namely $D_{F^2} \ge 1$ \cite{gammabound}.
Since $\beta'_{IR} > 0$, this bound is obviously satisfied.

% =======================================================================

\section{Calculation of $\beta'_{IR}$ to $O(\Delta_f^5)$ Order for 
${\rm SO}(4k+2)$ Theories} 
\label{betaprime_so_section}

For the SO($N$) theories with $N=4k+2$ considered here, namely SO(10), SO(14),
and SO(18) with $N_f$ fermions in the spinor representation, and $N_f$ in the
respective intervals in Table \ref{interval_so}, we calculate
\beq
d_2 = \frac{2^{N-1}}{3^2 (N-2)(11N^2+101N-224)} \ , 
\label{d2_so}
\eeq
\bigskip
\beq
d_3 = \frac{2^{(3N/2)-4}(3N^2+77N-160)}{3^3 (N-2)^2(11N^2+101N-224)^2} \ , 
\label{d3_so}
\eeq
\begin{widetext}
\beqs
d_4 &=& \frac{2^{2N-9}}{3^5 (N-2)^3 (11N^2+101N-224)^5} \, \bigg [
\Big ( -3993N^8+967780N^7-3621142N^6+40922980N^5 \cr\cr
&+&385439463N^4-5018429440N^3+18335731200N^2-28558381056N+16524705792 
\Big ) \cr\cr
&+&2^8 \cdot 33(11N^2+101N-224)\Big (11N^5-108N^4-1913N^3+17210N^2-50720N
+53376 \Big )\zeta_3 \ \bigg ] \ , 
\label{d4_so}
\eeqs
and
\beqs
d_5 &=& \frac{2^{(5N/2)-9}}{3^6 (N-2)^4(11N^2+101N-224)^7} \, \bigg [
\Big ( 464519N^{12}  -18008914N^{11} +359281505N^{10} -6749294188N^9 \cr\cr
&-&41411922215N^8 + 459185530094N^7 -1073251892065N^6 +3394219370864N^5 
-32099048433664N^4 \cr\cr
&+& 142779222543872N^3 -306826058932224N^2 + 326234208075776N -
138794015653888 \Big ) \cr\cr
&+&2^5(11N^2+101N-224)\Big (363N^{10} +38181N^9 + 1922118N^8 -35102518N^7 
-149165913N^6 +3972049185N^5 \cr\cr
&-&27149012488N^4 + 105670102816N^3 -249943359104N^2 +325769932800N 
-176231645184 \Big ) \zeta_3 \cr\cr
&-&2^7 \cdot 55(N-2)(11N^2+101N-224)^2\Big ( 33N^6 -27N^5 - 9221N^4 
+ 1879N^3 + 440008N^2 \cr\cr
&-& 2031648N + 2755584 \Big )\zeta_5 \ , 
\label{d5_so}
\eeqs
where $\zeta_s = \sum_{n=1}^\infty n^{-s}$ is the Riemann zeta function. 
Concerning the signs of these coefficients, $d_2$ and $d_3$ are manifestly 
positive (for a general $G$ and $R$) \cite{dex}, while the signs of $d_4$ and
$d_5$ depend on the theory. 

Evaluating these for the SO($N$) theories under consideration, we obtain
the following results for $\beta'_{IR}$ calculated up to $O(\Delta_f^5)$ order
(in floating-point format):
\beqs
&&{\rm SO}(10): \quad \beta'_{IR,\Delta_f^5} = 
 (3.7704725 \times 10^{-3})\Delta_f^2 
+(3.032105 \times 10^{-4})\Delta_f^3
-(1.2664165 \times 10^{-6})\Delta_f^4
-(5.4744784 \times 10^{-7})\Delta_f^5 \ , \cr\cr
&& {\rm where} \ \Delta_f = 22-N_f 
\label{betaprime_so10_p5}
\eeqs
\beqs
&&{\rm SO}(14): \quad \beta'_{IR,\Delta_f^5} =
 (2.266941 \times 10^{-2})\Delta_f^2 
+(4.534786 \times 10^{-3})\Delta_f^3 
+(2.0571128 \times 10^{-4})\Delta_f^4 
-(1.5915337 \times 10^{-5})\Delta_f^5 \ ,  \cr\cr
&& {\rm where} \ \Delta_f = 8.25-N_f 
\label{betaprime_so14_p5}
\eeqs
\beqs
&&{\rm SO}(18): \quad \beta'_{IR,\Delta_f^5} =
 0.176468 \Delta_f^2 
+0.100265 \Delta_f^3 
+(2.499877 \times 10^{-2})\Delta_f^4 
+(2.156910 \times 10^{-3})\Delta_f^5 \ , \cr\cr
&& {\rm where} \ \Delta_f = 2.75-N_f 
\label{betaprime_so18_p5}
\eeqs
\end{widetext}

\begin{figure}
  \begin{center}
    \includegraphics[height=6cm]{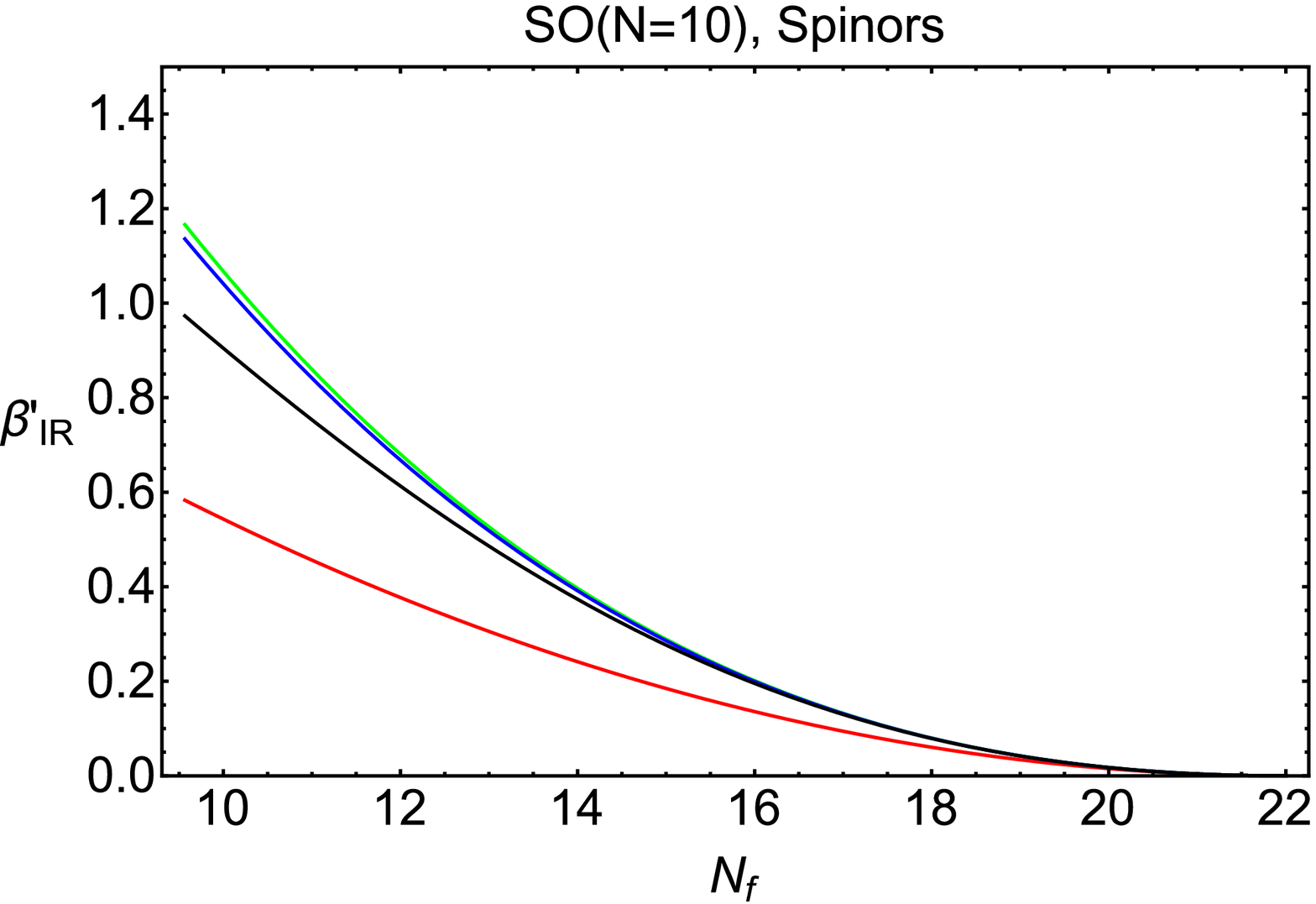}
  \end{center}
  \caption{Plot of $\beta'_{IR,\Delta_f^p}$ (labelled as
    $\beta'_{IR}$ on the vertical axis) for an SO(10) chiral gauge theory
    with fermions in the spinor representation, with $2 \le p \le 5$, 
    as a function of $N_f \in I$. At a given $N_f$, from bottom to top, the 
    curves (with colors online) refer to $\beta'_{IR,F,\Delta_f^2}$ (red),
    $\beta'_{IR,\Delta_f^5}$ (black), $\beta'_{IR,\Delta_f^4}$ (blue), and 
    $\beta'_{IR,\Delta_f^3}$ (green).}
\label{betaprime_SO10_plot}
\end{figure}

\begin{figure}
  \begin{center}
    \includegraphics[height=6cm]{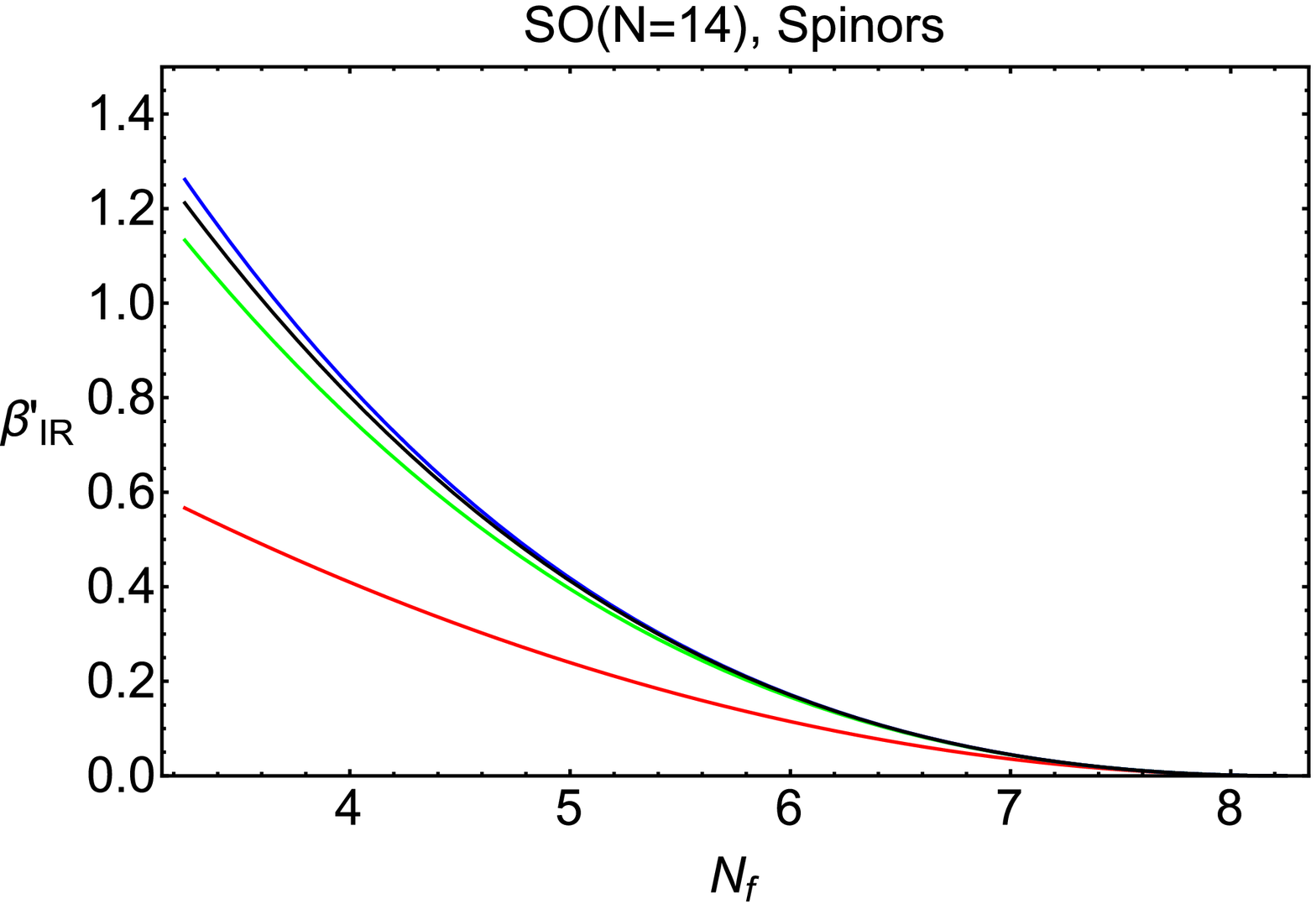}
  \end{center}
  \caption{Plot of $\beta'_{IR,\Delta_f^p}$ (labelled as
    $\beta'_{IR}$ on the vertical axis) for an SO(14) chiral gauge theory
    with fermions in the spinor representation, with $2 \le p \le 5$, 
    as a function of $N_f \in I$. At a given $N_f$, from bottom to top, the 
    curves (with colors online) refer to $\beta'_{IR,F,\Delta_f^2}$ (red),
    $\beta'_{IR,\Delta_f^3}$ (green), $\beta'_{IR,\Delta_f^5}$ (black), and
    $\beta'_{IR,\Delta_f^4}$ (blue).}
\label{betaprime_SO14_plot}
\end{figure}

\begin{figure}
  \begin{center}
    \includegraphics[height=6cm]{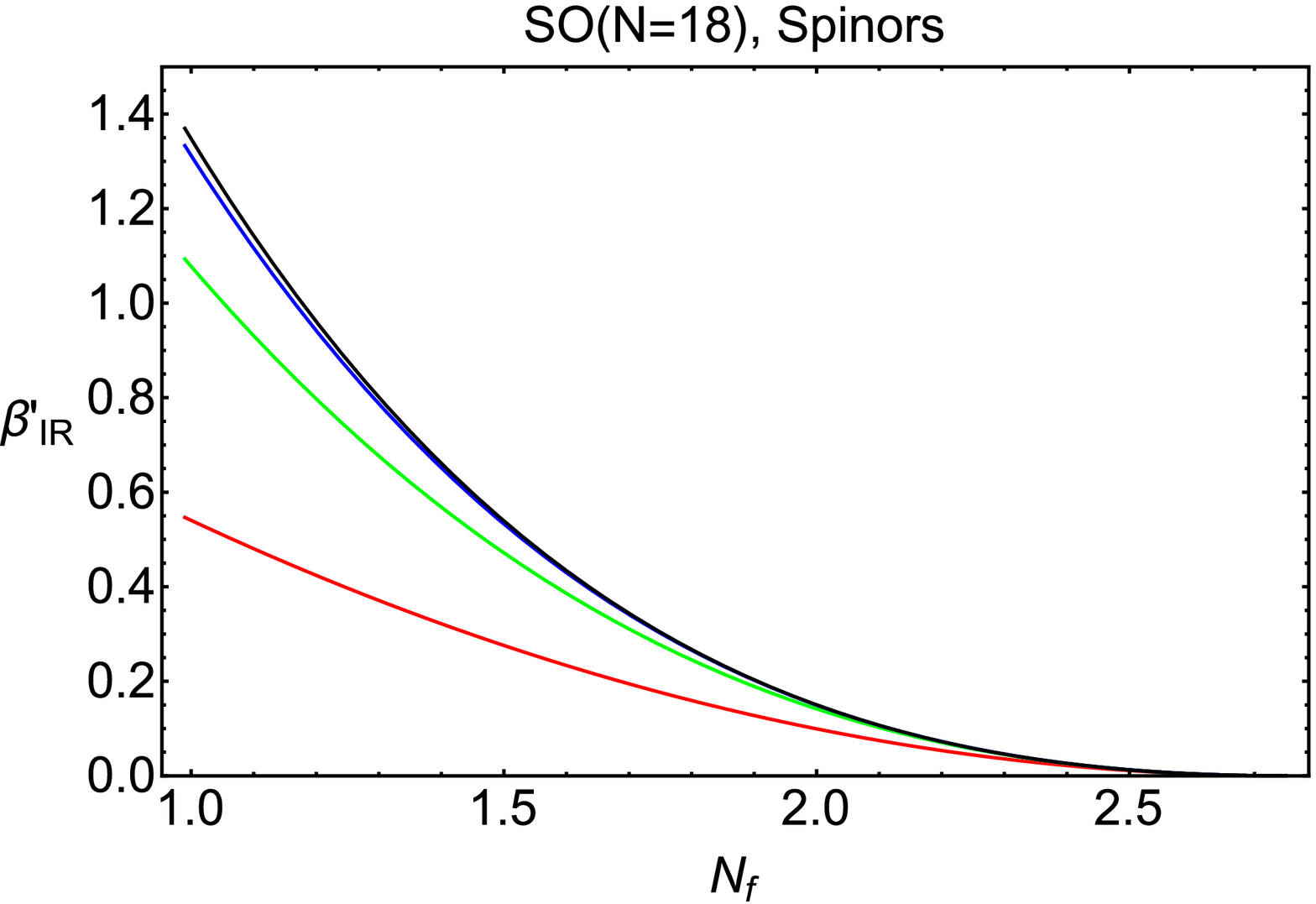}
  \end{center}
  \caption{Plot of $\beta'_{IR,\Delta_f^p}$ (labelled as
    $\beta'_{IR}$ on the vertical axis) for an SO(18) chiral gauge theory
    with fermions in the spinor representation, with $2 \le p \le 5$, 
    as a function of $N_f \in I$. At a given $N_f$, from bottom to top, the 
    curves (with colors online) refer to $\beta'_{IR,F,\Delta_f^2}$ (red),
    $\beta'_{IR,\Delta_f^3}$ (green), $\beta'_{IR,\Delta_f^4}$ (blue), and 
    $\beta'_{IR,\Delta_f^5}$ (black).}
\label{betaprime_SO18_plot}
\end{figure}

In Figs. \ref{betaprime_SO10_plot}-\ref{betaprime_SO18_plot} we plot the
resultant values of $\beta'_{IR,\Delta_f^p}$ with $2 \le p \le 5$ for these
theories. 

% =====================================================================

\section{Calculation of $\beta'_{IR}$ to $O(\Delta_f^5)$ Order for 
E$_6$ Theory} 
\label{betaprime_e6_section}

\begin{figure}
  \begin{center}
    \includegraphics[height=6cm]{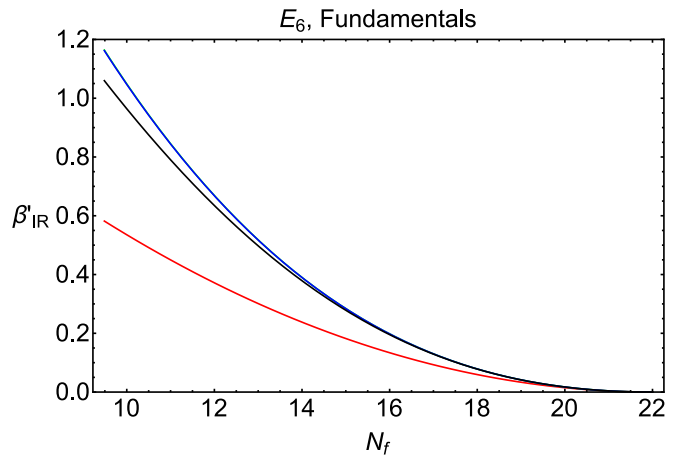}
  \end{center}
  \caption{Plot of $\beta'_{IR,\Delta_f^p}$ (labelled as
    $\beta'_{IR}$ on the vertical axis) for an E$_6$ chiral gauge theory
    with fermions in the fundamental representation, with $2 \le p \le 5$, 
    as a function of $N_f \in I$. At a given $N_f$, from bottom to top, the 
    curves (with colors online) refer to $\beta'_{IR,F,\Delta_f^2}$ (red),
    $\beta'_{IR,\Delta_f^5}$ (black), $\beta'_{IR,\Delta_f^4}$ (blue), and
    $\beta'_{IR,\Delta_f^3}$ (green).  Note that the curves for $p=3$ and
    $p=4$ are too close to each other to be distinguished in the plot.}
\label{betaprime_E6_plot}
\end{figure}

For the E$_6$ theory with $N_f$ fermions in the fundamental (27-dimensional)
representation, we calculate 
\beq
d_2 = \frac{1}{269} = 3.717472 \times 10^{-3} \ , 
\label{d2_e6}
\eeq
\beq
d_3 = \frac{43}{2 \cdot (269)^2} = 2.971214 \times 10^{-4}  \ , 
\label{d3_e6}
\eeq
\beqs
d_4 &=& \frac{660297341}{2^3 \cdot 3^2 \cdot (269)^5} 
-\frac{2 \cdot 14355}{(269)^4}\zeta_3 \cr\cr
    &=& -(0.7999706 \times 10^{-7})  \ , 
\label{d4_e6}
\eeqs
and
\beqs
d_5 &=& -\frac{328284821696663}{2^5 \cdot 3^4 \cdot (269)^7} 
-\frac{2^8 \cdot 18928393}{3^3 \cdot (269)^6}\zeta_3 \cr\cr
&+&\frac{2^3 \cdot 251075}{(269)^5}\zeta_5 = -(3.3333007 \times 10^{-7}) \ . 
\cr\cr
&&
\label{d5_e6}
\eeqs
Hence, to $O(\Delta_f^5)$, with $\Delta_f=22-N_f$ here 
(in floating-point format), 
\beqs
\beta'_{IR,\Delta_f^5} &&= (3.717472 \times 10^{-3})\Delta_f^2 + 
(2.971214 \times 10^{-4})\Delta_f^3 \cr\cr
&&-(0.7999706 \times 10^{-7}) \Delta_f^4 
-(3.3333007 \times 10^{-7}) \Delta_f^5 \ . \cr\cr
&&
\label{betaprime_e6_fund}
\eeqs
In Fig. \ref{betaprime_E6_plot} we plot the
resultant values of $\beta'_{IR,\Delta_f^p}$ with $2 \le p \le 5$ for this
E$_6$ theory.  Because $|d_4| \ll d_3$, the curve for $\beta'_{IR,\Delta_f^4}$ 
is too close to the curve for $\beta'_{IR,\Delta_f^3}$ to be distinguished from
it in the plot.

% =======================================================================

\section{Conclusions}
\label{conclusions_section}

In conclusion, in this paper we have presented scheme-independent 
calculations of $\beta'_{IR}$ at an IR fixed point in the non-Abelian Coulomb
phase of several asymptotically 
free chiral gauge theories, namely SO($4k+2$) with $2 \le k \le 4$, with 
respective numbers $N_f$ of fermions in the spinor representation, and E$_6$ 
with fermions in the fundamental representation.  Our results further elucidate
the properties of (four-dimensional) conformal field theories and should serve
as useful inputs for the resurgent research on these theories. 

% =======================================================================

\begin{acknowledgments}

The research of T.A.R. and R.S. was supported in part by the Danish National
Research Foundation grant DNRF90 to CP$^3$-Origins at SDU and by the
U.S. National Science Foundation Grant NSF-PHY-16-1620628, respectively.  

\end{acknowledgments}

% =======================================================================

\begin{appendix}

\section{Some Group-Theoretic Quantities}
\label{group_invariants}

\bigskip

In this appendix we discuss some relevant group-theoretic quantities.  The
generators of the Lie algebra of $G$ in the representation $R$ are denoted
$T^a_R$, where $a$ is a group index. These satisfy $[T^a_R,T^b_R]=if^{abc}
T^c_R$ (where sums over repeated group indices are understood). We denote the
dimension of a given representation $R$ as $d_R = {\rm dim}(R)$, and denote $A$
as the adjoint representation. The trace invariant is defined by ${\rm
  Tr}_R(T^a_R T^b_R) = T(R)\delta_{ab}$ and the quadratic Casimir invariant
$C_2(R)$ is given by $T^a_RT^a_R = C_2(R) I$, where $I$ is the $d_R \times d_R$
identity matrix.  For a fermion $f$ in $R$, a compact notation is $T_f \equiv
T(R)$, $C_f \equiv C_2(R)$, and $C_A \equiv C_2(A)$. As discussed in
\cite{dexo}, although these group invariants depend on a convention for the
normalization of the structure constants $f^{abc}$, the $d_j$ are independent
of this convention.

The general expressions for the coefficients $d_4$ and $d_5$ 
\cite{dex,dexl} involve certain quartic group invariants \cite{quartic}. 
For SO($N$) with $N=4k+2$, we calculate these to be 
\beqs
&& {\rm SO}(N), \ R = {\rm spinor}: \cr\cr
   && \frac{d_R^{abcd}d_A^{abcd}}{d_A} = 
-\frac{2^{(N/2)-8}(N-2)(N^2-22N+52)}{3}, \cr\cr
   && \frac{d_R^{abcd}d_R^{abcd}}{d_A} = 
\frac{2^{N-15}(13N^2-61N+76)}{3} \ . 
\label{quartics_so}
\eeqs
We gave the quartic invariant $d_A^{abcd}d_A^{abcd}/d_A$ for SO($N$) 
previously in \cite{dexo}; for reference, it is 
\beq
\frac{d_A^{abcd}d_A^{abcd}}{d_A} = \frac{(N-2)(N^3-15N^2+138N-296)}{24} \ . 
\label{dada_over_na_so}
\eeq
For E$_6$ with $R=F$, the fundamental representation, we calculate 
\beqs
{\rm E}_6: && \quad \frac{d_A^{abcd}d_A^{abcd}}{d_A} = 540, \qquad 
\frac{d_F^{abcd}d_A^{abcd}}{d_A} = 90, \cr\cr
&& \frac{d_F^{abcd}d_F^{abcd}}{d_A} = 15 \ . 
\label{quartics_e6}
\eeqs

\end{appendix} 

% ========================================================================

% =========================================================================

\newpage

\begin{table}
  \caption{\footnotesize{Interval $I$ in terms of $N_f$, for 
for $N_f$ formally generalized to real numbers, ${\mathbb R}_+$ and for 
physical, integral values of $N_f \in {\mathbb N}_+$, 
    for the $G={\rm SO}(4k+2)$ chiral gauge theories with $k=2, \ 3, \ 4$, 
   i.e., SO(10), SO(14), and SO(18) and chiral fermions in the spinor
      representation.}}
\begin{center}
\begin{tabular}{|c||c|c|c|} \hline\hline
$G$ & $I$, \ $N_f \in {\mathbb R}_+$ & $I$, $N_f \in {\mathbb N}_+$ \\
\hline 
SO(10) & $9.565 < N_f < 22$ & $10 \le N_f \le 21$ \\
\hline
SO(14) & $3.251 < N_f < 8.25$ & $4 \le N_f \le 8$ \\
\hline
SO(18) & $0.990 < N_f < 2.75$ & $1 \le N_f \le 2$ \\
\hline\hline
\end{tabular}
\end{center}
\label{interval_so}
\end{table}

\end{document}